\documentclass[iop]{emulateapj}  

\usepackage{amsmath}

\newcommand\kms{\ifmmode{\rm km\thinspace s^{-1}}\else km\thinspace s$^{-1}$\fi}
\newcommand\vstar{80\thinspace Tau}
\newcommand\hip{{\it Hipparcos\/}}
\newcommand\gaia{{\it Gaia\/}}

\shortauthors{Torres et al.}
\shorttitle{80 Tau}

\begin{document} 

\submitted{Accepted for publication in The Astrophysical Journal}

\title{Dynamical masses for the Hyades binary 80 Tauri}

\author{
Guillermo Torres
}

\affiliation{Center for Astrophysics $\vert$ Harvard \&
  Smithsonian, 60 Garden St., Cambridge, MA 02138, USA;
  gtorres@cfa.harvard.edu}

\begin{abstract}

The empirical mass-luminosity relation in the Hyades cluster rests on
dynamical mass determinations for five binary systems, of which one is
eclipsing and the other four are visual or interferometric binaries.
The last one was identified and first measured more than 20 years
ago. Here we present dynamical mass measurements for a new binary
system in the cluster, \vstar, which is also a visual system with a
much longer orbital period of about 170~yr.  Although we lack the
radial-velocity information that has enabled the individual mass
determinations in all of the previous binaries, we show that it is
still possible to derive the component masses for \vstar\ using only
astrometric observations. This is enabled by the accurate proper
motion measurements from the \hip\ and \gaia\ missions, which
constrain the orbital acceleration in the plane of the sky. Separate
proper motion values from \gaia\ for the primary and secondary provide
a direct constraint on the mass ratio. Our mass measurements, $M_{\rm
  A} = 1.63^{+0.30}_{-0.13}~M_{\sun}$ and $M_{\rm B} =
1.11^{+0.21}_{-0.14}~M_{\sun}$, are consistent with the
mass-luminosity relation defined by the five previously known systems,
which in turn is in good agreement with current models of stellar
evolution.

\end{abstract}


\section{Introduction}
\label{sec:introduction}

The empirical mass-luminosity relation (MLR) in the Hyades cluster has
been the subject of numerous investigations over the past
decades. This mapping between the mass of a star ---its most
fundamental property--- and its brightness offers a valuable way to
test models of stellar evolution in a homogeneous population with a
well-known age \citep[625~Myr;][]{Perryman:1998} and chemical
composition \citep[${\rm [Fe/H]} = +0.18 \pm
  0.03$;][]{Dutra-Ferreira:2016}.  Progress toward building the Hyades
MLR has been relatively slow, however, as the essential ingredients
are the dynamical masses, and only a handful of binary systems
suitable for this type of determination have been found so far, and
measured sufficiently well.

The first example was the 5.6-day eclipsing binary V818\thinspace Tau
(HD\thinspace 27130 = vB\thinspace 22)\footnote{We provide for
  reference the vB designations originating with
  \cite{vanBueren:1952}, which are also in common use in the
  literature.}, discovered by \cite{McClure:1982}, who published
preliminary determinations of the individual masses that were
subsequently refined by \cite{Schiller:1987}. This remains the only
known eclipsing binary in the cluster consisting of normal
stars. Dynamical masses for a second system, the
spectroscopic-interferometric binary $\theta^2$\thinspace Tau
(HD\thinspace 28139 = vB\thinspace 72, $P = 141$\thinspace days), were
reported by \cite{Peterson:1993} and later revised by
\cite{Tomkin:1995} to remove an assumption on the mass ratio. Next
came the long-period visual binaries 51\thinspace Tau (HD\thinspace
27176 = vB\thinspace 24, $P = 11.3$~yr), 70\thinspace Tau
(HD\thinspace 27991 = vB\thinspace 57, $P = 6.3$~yr), and
$\theta^1$\thinspace Tau (HD\thinspace 28307 = vB\thinspace 71, $P =
16.3$~yr), studied, respectively, by \cite{Torres:1997a},
\cite{Torres:1997b}, and \cite{Torres:1997c}. The primary of the
latter binary is one of the four giants in the Hyades. All five
systems have been revisited over the years by other authors with
improvements in the dynamical masses in some cases, or in the absolute
magnitudes, but no other binaries that are amenable to
model-independent mass determinations for the individual components
have been found in more than 20 years.

In this paper we report the first dynamical mass measurements for a
new system, 80\thinspace Tau (HD\thinspace 28485 = vB\thinspace 80 =
ADS\thinspace 3264), a bright ($V = 5.55$), $\sim$170~yr visual binary
discovered by \cite{Struve:1837}. Its membership in the Hyades is well
established \citep{vanBueren:1952}, and is supported by results from
the \gaia\ mission, among many others.  While the five previous
systems have relied in part on spectroscopy (radial velocities) to
infer the mass of each star, for \vstar\ this is not practical because
of the long orbital period and the fact that the primary is a very
rapid rotator ($v \sin i \approx 180~\kms$).  This hinders the
determination of velocities with the requisite precision.
Nevertheless, as we show below, it is still possible to derive the
individual masses using only astrometry. This is enabled by two types
of complementary astrometric observations: \emph{i)} ground-based
measurements of the relative positions of the components collected by
dozens of visual observers over a full cycle of the orbit; and
\emph{ii)} accurate parallaxes and proper motion (p.m.)\ measurements
for both stars provided by the \gaia\ mission, and for the primary by
\hip.  Together these p.m.\ constrain the mass ratio and contain
valuable information on astrometric accelerations, by virtue of the
24-yr interval between the two space missions.

We describe the astrometric observations in
Section~\ref{sec:astrometry}. Our modeling of the orbital motion is
discussed in Section~\ref{sec:analysis}, where we also present the
results.  We then use them in Section~\ref{sec:MLR} to investigate the
MLR in the Hyades and compare the empirical relation with current
models of stellar evolution. Final remarks are given in
Section~\ref{sec:remarks}.

\section{Astrometric observations}
\label{sec:astrometry}

\subsection{Visual measurements}
\label{sec:visual}

The discovery of \vstar\ as a visual binary is credited to
\cite{Struve:1837}, who first measured it in 1831 finding a companion
some 2.5 magnitudes fainter at a distance of 1\farcs74 due north. In
his honor the object also carries the double star designation
STF\thinspace 554. Some 180 measurements of the position angle and
angular separation ($\theta$, $\rho$) have been recorded in the 180
years since, the latest one in 2015. They cover a full orbital cycle,
with a periastron passage around 1892.  The vast majority of these
observations were made with a classical filar micrometer, while a few
of the more recent ones used the speckle interferometry technique or
adaptive optics imaging.  A listing of these measurements as contained
in the Washington Double Star Catalog (WDS) was kindly provided by
Brian Mason (U.S.\ Naval Observatory, Washington D.C.), with the dates
of observation uniformly transformed from Besselian years to Julian
years. To these we have added one observation from 1897 that was not
included in the WDS listing, which was made by \cite{Burnham:1906}
very close to periastron, and one from the \gaia/DR2 Catalogue
\citep{Gaia:2018} that we constructed by differencing the positions of
the two components measured separately by the mission.

As a result of its long observational history, several visual orbits
of \vstar\ have been computed over the years, and it was recognized
early on as being a very eccentric and highly inclined system, such
that all of the measurements are in the first quadrant.  The orbit by
\cite{Baize:1980}, based on observations up to about 1975, is the one
featured in the Sixth Catalog of Orbits of Visual Binary
Stars\footnote{\url{https://ad.usno.navy.mil/wds/orb6.html}}
maintained at the U.S.\ Naval Observatory. According to
\cite{Baize:1980} it has a period of 180~yr, a semimajor axis of
1\farcs00, an eccentricity of 0.82, and an inclination of
107\fdg6. However, that orbit no longer matches the more recent data.
A revised model was proposed by \cite{Peterson:1988} that is more
eccentric ($e = 0.901 \pm 0.027$), and a more recent one by
\cite{Izmailov:2019} increases the eccentricity even further to $0.979
\pm 0.011$.

While a few of the recent visual observations of \vstar\ were
published with corresponding uncertainties, most of them have no
reported uncertainties at all.  Because the quality of any particular
measurement depends sensitively on many factors including the aperture
used, the observing conditions, and even the experience and
disposition of the observer, the task of assigning realistic
uncertainties to observations of this kind is fraught with difficulty,
and there is no unique way to do this.  After some experimentation we
have chosen to separate the observations into four groups, and to
assign nominal uncertainties to the angular separations and position
angles within each group. For the position angles the uncertainties
were expressed in seconds of arc in the tangential direction,
$\sigma_{\rm t}$, to account explicitly for the dependence of the
angular precision on angular separation: $\sigma_{\rm t} =
\rho\thinspace \sigma_{\theta}$. This is important to properly weight
the very few observations made near periastron, which were especially
challenging for this binary given the telescopes in use at the time,
as noted by \cite{Baize:1980}. The four groups we considered are: 1)
observations prior to 1900, which were all made with filar micrometers
and small aperture telescopes; these were assigned errors of
$\sigma_{\rho} = 0\farcs13$ and $\sigma_{\rm t} = 0\farcs10$; 2)
micrometer observations from 1900 to about 1975, which received
$\sigma_{\rho}$ and $\sigma_{\rm t}$ errors of 0\farcs06 and
0\farcs025, respectively; 3) more recent micrometer measurements
($\sigma_{\rho} = 0\farcs03$ and $\sigma_{\rm t} = 0\farcs042$); and
4) speckle measurements without published uncertainties
($\sigma_{\rho} = 0\farcs014$ and $\sigma_{\rm t} =
0\farcs012$). These values of $\sigma_{\rho}$ and $\sigma_{\rm t}$ are
the result of iterations of the analysis described later in
Section~\ref{sec:analysis} intended to achieve balanced residuals
among the different groups, i.e., to establish the \emph{relative
  weighting} among them. Further adjustments to the \emph{absolute
  scale} of the separation and position angle errors are described
below.

\setlength{\tabcolsep}{6pt}  
\begin{deluxetable*}{lcccccccc}
\tablewidth{0pc}
\tablecaption{Visual Observations of \vstar \label{tab:visual}}
\tablehead{
\colhead{Year} &
\colhead{$\theta$ (\arcdeg)} &
\colhead{$\sigma_{\rm t}$ (\arcsec)} &
\colhead{$(O-C)_{\theta}$ (\arcdeg)} &
\colhead{$(O-C)_{\theta}/\sigma_{\theta}$} &
\colhead{$\rho$ (\arcsec)} &
\colhead{$\sigma_{\rho}$ (\arcsec)} &
\colhead{$(O-C)_{\rho}$ (\arcsec)} &
\colhead{$(O-C)_{\rho}/\sigma_{\rho}$}
}
\startdata
 1831.18  &   12.9  &  0.146  & $-$3.68  & $-$0.71 &  1.74    &  0.193 & \phs0.139 & \phs0.72 \\
 1836.96  & \phn9.8 &  0.146  & $-$6.14  & $-$1.15 &  1.5\phn &  0.193 & $-$0.059  & $-$0.30 \\
 1837.22  &   11.0  &  0.146  & $-$4.91  & $-$0.92 &  1.4\phn &  0.193 & $-$0.156  & $-$0.81 \\
 1839.16  &   13.9  &  0.146  & $-$1.79  & $-$0.33 &  1.6\phn &  0.193 & \phs0.061 & \phs0.31 \\
 1840.12  &   14.8  &  0.146  & $-$0.78  & $-$0.14 &  1.66    &  0.193 & \phs0.130 & \phs0.67     
\enddata

\tablecomments{The position angles listed are the original ones;
  precession corrections were applied internally during the orbital
  analysis. The uncertainties listed in the table correspond to the
  $\sigma_{\rho}$ and $\sigma_{\rm t}$ values given in the text
  multiplied by the scaling factors $f_{\rho}$ and $f_{\theta}$
  reported in Section~\ref{sec:analysis}, and represent the final
  errors used in the solution described there. We also list the
  residuals in angular separation and position angle normalized to
  their uncertainties $\sigma_{\rho}$ and $\sigma_{\theta} =
  \sigma_{\rm t}/\rho$. (This table is available in its entirety in
  machine-readable form.)}

\end{deluxetable*}
\setlength{\tabcolsep}{6pt}  

Two measurements of \vstar\ that were flagged as erroneous in the WDS
were rejected, along with one position angle and three angular
separation measurements that we found gave anomalously large residuals
and were very different from others near in time. We also reversed the
quadrant of an observation from 1886.97 that had previously been
reversed in the WDS, as we found that the original angle as published
fits our new orbit well. Finally, precession corrections were applied
to all position angles to reduce them to the year 2000.0. In all we
used 177 measures of the angular separation and 183 of the position
angle. We list all observations in Table~\ref{tab:visual} along with
their uncertainties and residuals from our adopted model orbit
discussed below.

\subsection{\hip\ and \gaia}
\label{sec:propermotions}

\vstar\ is listed in both the \hip\ Catalogue
\citep[HIP~20995;][]{vanLeeuwen:2007} and the \gaia/DR2 Catalogue
\citep{Gaia:2018}, the latter containing separate entries for the
primary and secondary.\footnote{DR2~3312536927686011520 and
  DR2~3312536923393893120, respectively.}  The two \gaia\ estimates of
the trigonometric parallax are consistent with each other, and with
the \hip\ value for the primary.  Knowledge of the distance along with
the visual orbit can then immediately provide a measure of the total
mass of the system.

Beyond the visual observations described above, other constraints on
the relative orbit that are available include the proper motions
measured by the two missions, which are different. The mean epochs of
those measurements are $\sim$1991.25 for \hip\ and $\sim$2015.5 for
\gaia. This 24-yr interval is long enough that orbital motion affects
the p.m.\ determination at each epoch, providing a measure of the
acceleration in the plane of the sky.  \cite{Brandt:2018} performed a
cross-calibration of the \hip\ and \gaia\ catalogs to place them on
the reference frame of \gaia/DR2, and produced a catalog of
accelerations consisting of three nearly independent proper motions
for each star: one p.m.\ near each mean epoch ($\mu_{\rm H}$ and
$\mu_{\rm G}$ for \hip\ and \gaia, with components in R.A.\ and Dec.),
and a third measurement given by the \gaia--\hip\ positional
difference divided by the 24-yr time baseline ($\mu_{\rm HG}$). As
this last quantity is usually much more precise than the other two,
\cite{Brandt:2018} advocated subtracting $\mu_{\rm HG}$ from both
$\mu_{\rm H}$ and $\mu_{\rm G}$, which cancels out the motion of the
barycenter (typically irrelevant for mass determinations). This gives
two p.m.\ differences that are well suited for constraining the
relative orbit of a binary (i.e., the total mass), even in cases where
other types of observations cover only a fraction of the orbit.
Examples of the use of this astrometric information combined with
visual observations and with radial-velocity measurements to infer
component masses for binaries were given by \cite{Dupuy:2019} and
\cite{Brandt:2019}.

In those studies the radial velocities (typically in the form of a
measured velocity change for the primary) served to provide the
necessary constraint on the mass ratio, which then enabled the
individual masses to be calculated. For \vstar\ we unfortunately lack
sufficiently precise measurements of the radial velocities of the
components to show a statistically significant change with time. This
is due to the fact that the orbital period is very long and we happen
to be far from periastron, so that the stars are now moving very
slowly at about the same speed when projected along the line of sight,
i.e., the spectral lines are completely blended. Furthermore, the
primary star is a rapid rotator \citep[$v \sin i =
  180~\kms$;][]{Royer:2002}, making precise velocity measurements
extraordinarily difficult.\footnote{The many existing velocity
  measurements of this bright star in the literature show considerable
  scatter most likely due to this difficulty \citep{Abt:1980}. A claim
  has been made of periodic variability with an orbital period of 30.5
  days and a center-of-mass velocity of $\gamma = +29.3~\kms$
  \citep{Heintz:1981}. This seems dubious, however, as $\gamma$ is
  more than 10~\kms\ lower than the velocity expected for \vstar\ from
  its position in the cluster, and especially since more recent
  velocity measurements have failed to follow that orbit.}  Thus, we
are missing the spectroscopic constraint on the mass ratio.

However, because \gaia\ measured the proper motion of both components
($\mu_{\rm G,A}$, $\mu_{\rm G,B}$), and the values are appreciably
different, there is astrometric information on the mass ratio that
comes from the fact that the orbits of the two stars around their
common center of mass perturb the p.m.\ measured by \gaia\ in
different ways that are directly related to their masses.
Conceptually, then, provided the p.m.\ of the center of mass ($\mu_0$)
can be determined, the ratio between the differences \mbox{$\mu_{\rm
    G,A}-\mu_0$} and \mbox{$\mu_{\rm G,B}-\mu_0$} yields the mass
ratio $M_{\rm B}/M_{\rm A}$. This is analogous to the situation in the
direction perpendicular to the plane of the sky, in which a single
measurement of the primary and secondary radial velocity in a binary
at a given time is sufficient to establish the mass ratio if the
center-of-mass velocity is known. The accuracy of the mass ratio
determination from the proper motions will depend critically on how
well the motion of the barycenter can be constrained by other
measurements.

\setlength{\tabcolsep}{6pt}  
\begin{deluxetable*}{lcccccc}
\tablewidth{0pc}
\tablecaption{Proper Motion and Parallax Information for \vstar\ from
  \gaia/DR2 and \hip \label{tab:pm}}
\tablehead{
\colhead{Source} &
\colhead{Comp} &
\colhead{$\mu_{\alpha}^*$ (mas yr$^{-1}$)} &
\colhead{$\mu_{\delta}$ (mas yr$^{-1}$)} &
\colhead{Corr} &
\colhead{Average Epoch} &
\colhead{$\pi$ (mas)}
}
\startdata
HIP        & A & $+107.33 \pm 1.05$   & $-23.56 \pm 0.64$   & $-0.090$ & 1991.35 / 1991.04 & $22.26 \pm 1.01$ \\
\gaia--HIP & A & $+108.446 \pm 0.034$ & $-23.719 \pm 0.021$ & $+0.244$ & \nodata & \nodata \\
\gaia      & A & $+108.92 \pm 0.43$   & $-22.14 \pm 0.27$   & $-0.293$ & 2015.58 / 2015.63 & $21.12 \pm 0.12$ \\
\gaia      & B & $+102.57 \pm 0.30$   & $-31.43 \pm 0.15$   & $-0.314$ & 2015.58 / 2015.63 & $20.934 \pm 0.069$
\enddata
\tablecomments{The first three entries are taken from the acceleration
  catalog of \cite{Brandt:2018}; the last is from \gaia/DR2. ``Comp''
  refers to the primary or secondary component, $\mu_{\alpha}^*$
  represents the p.m.\ in R.A.\ multiplied by the cosine of the
  Declination, and ``Corr'' is the correlation coefficient between the
  proper motions in R.A.\ and Dec. The ``Average Epoch'' is given
  separately for the p.m.\ measurements in R.A. and Dec. The
  \hip\ parallax was calculated from the values reported in the
  revised and original versions of the catalog \citep{ESA:1997,
    vanLeeuwen:2007} with the same 60/40 linear combination
  recommended by \cite{Brandt:2018}, following his Eq.(18).}
\end{deluxetable*}
\setlength{\tabcolsep}{6pt}  

To implement this idea we therefore proceeded differently than
advocated by \cite{Brandt:2018}: instead of using the differences
$\mu_{\rm H}-\mu_{\rm HG}$ and $\mu_{\rm G}-\mu_{\rm HG}$ for the
primary as our measurements to supplement the constraint on the
relative orbit provided by the visual observations, we used the four
p.m.\ measurements $\mu_{\rm H}$, $\mu_{\rm HG}$, $\mu_{\rm G,A}$ and
$\mu_{\rm G,B}$ as independent observables to constrain the orbits of
both stars (and therefore their mass ratio), and to solve for $\mu_0$
at the same time.  These four measurements are listed in
Table~\ref{tab:pm} along with the correlation coefficients between the
R.A.\ and Dec.\ components as given by \cite{Brandt:2018} or the
\gaia\ Catalogue. We also give the mean epoch in each coordinate, as
well as the parallaxes. The \hip\ parallax we report is the result of
combining the values from the revised and original versions of the
catalog \citep{ESA:1997, vanLeeuwen:2007} in the same 60/40
proportions as recommended by \cite{Brandt:2018}, adding the same
0.20~mas error inflation in quadrature as prescribed by his
Eq.(18). The uncertainty for the \gaia\ p.m.\ of the secondary has
also been adjusted upward as prescribed by \cite{Brandt:2018}.

\section{Analysis and Results}
\label{sec:analysis}

Our procedure combines the visual observations with the four
p.m.\ measurements from \hip\ and \gaia\ into a single solution.  The
visual orbit is described by the orbital period ($P$), the semimajor
axis ($a$), the eccentricity parameters $\sqrt{e}\cos\omega_{\rm B}$
and $\sqrt{e}\sin\omega_{\rm B}$ (where $e$ is the eccentricity and
$\omega_{\rm B}$ the longitude of periastron for the secondary), the
cosine of the orbital inclination angle ($\cos i$), the position angle
of the ascending node\footnote{In the absence of radial velocity
  information we followed the usual convention of assuming the node
  for which $\Omega < 180\arcdeg$ to be the ascending node.} for the
equinox J2000 ($\Omega$), and a reference time of periastron passage
($T$).  Additionally we solved for the mass fraction $f = M_{\rm
  B}/(M_{\rm A}+M_{\rm B})$ (more convenient for our purposes than the
mass ratio) and the R.A.\ and Dec.\ components of the proper motion of
the barycenter, $\mu^*_{\alpha,0}$ and $\mu_{\delta,0}$. The notation
$\mu^*_{\alpha}$ we use here and in the following represents the
p.m.\ in R.A.\ multiplied by the cosine of the Declination.
 
Our method of solution used the {\tt emcee\/}\footnote{\url
  https://emcee.readthedocs.io/en/v2.2.1/} code of
\cite{Foreman-Mackey:2013}, which is a Python implementation of the
affine-invariant Markov Chain Monte Carlo (MCMC) ensemble sampler
proposed by \cite{Goodman:2010}. We used 100 walkers with 20,000 links
each, after discarding the burn-in, and priors for all variables were
assumed to be uniform except for $P$ and $a$, for which we used
log-uniform priors. To establish the optimal relative weighting
between the angular separation and position angle measurements we
included two more parameters, $f_{\rho}$ and $f_{\theta}$, which
represent multiplicative scaling factors for the uncertainties
described in Section~\ref{sec:visual}. They are intended to result in
reduced $\chi^2$ values near unity for each type of observation, and
were solved for simultaneously and self-consistently with the other
parameters \citep[see][]{Gregory:2005}. The adopted priors for
$f_{\rho}$ and $f_{\theta}$ were log-uniform.  Convergence of our MCMC
procedure was checked by visual inspection of the chains, and by
requiring a Gelman-Rubin statistic \citep{Gelman:1992} less than 1.05
for all 12 adjustable parameters. Our likelihood function is
$$\ln \mathcal{L} = -0.5(\chi^2_{\rho} + \chi^2_{\theta} +
\chi^2_{\rm H} + \chi^2_{\rm HG} + \chi^2_{\rm G,A} + \chi^2_{\rm G,B}),$$
where the chi-squared terms in this expression are
\begin{align}
\chi^2_{\rho} &= \sum_{i=1}^{N_{\rho}} \frac{(\rho_i-\hat\rho[t_i])^2}{(f_{\rho} \sigma_{\rho})^2} +
\sum_{i=1}^{N_{\rho}} \ln (f_{\rho} \sigma_{\rho})^2 \nonumber \\
\chi^2_{\theta} &= \sum_{i=1}^{N_{\theta}} \frac{(\theta_i-\hat\theta[t_i])^2}{(f_{\theta} \sigma_{\theta})^2} +
\sum_{i=1}^{N_{\theta}} \ln (f_{\theta} \sigma_{\theta})^2 \nonumber \\
\chi^2_{\rm H} &= (\mu^*_{\alpha,{\rm H}}-\hat\mu^*_{\alpha,{\rm H}})^2 C^{-1}_{\alpha\alpha,{\rm H}} +
(\mu_{\delta,{\rm H}}-\hat\mu_{\delta,{\rm H}})^2 C^{-1}_{\delta\delta,{\rm H}} \nonumber \\
& + 2(\mu^*_{\alpha,{\rm H}}-\hat\mu^*_{\alpha,{\rm H}})(\mu_{\delta,{\rm H}}-\hat\mu_{\delta,{\rm H}})
C^{-1}_{\alpha\delta,{\rm H}} \nonumber \\
\chi^2_{\rm HG} &= (\mu^*_{\alpha,{\rm HG}}-\hat\mu^*_{\alpha,{\rm HG}})^2 C^{-1}_{\alpha\alpha,{\rm HG}} \nonumber \\
& + (\mu_{\delta,{\rm HG}}-\hat\mu_{\delta,{\rm HG}})^2 C^{-1}_{\delta\delta,{\rm HG}} \nonumber \\
& + 2(\mu^*_{\alpha,{\rm HG}}-\hat\mu^*_{\alpha,{\rm HG}})(\mu_{\delta,{\rm HG}}-\hat\mu_{\delta,{\rm HG}})
C^{-1}_{\alpha\delta,{\rm HG}} \nonumber \\
\chi^2_{\rm G,A} &= (\mu^*_{\alpha,{\rm G,A}}-\hat\mu^*_{\alpha,{\rm G,A}})^2 C^{-1}_{\alpha\alpha,{\rm G,A}} \nonumber \\
& + (\mu_{\delta,{\rm G,A}}-\hat\mu_{\delta,{\rm G,A}})^2 C^{-1}_{\delta\delta,{\rm G,A}} \nonumber \\
& + 2(\mu^*_{\alpha,{\rm G,A}}-\hat\mu^*_{\alpha,{\rm G,A}})(\mu_{\delta,{\rm G,A}}-\hat\mu_{\delta,{\rm G,A}})
C^{-1}_{\alpha\delta,{\rm G,A}} \nonumber \\
\chi^2_{\rm G,B} &= (\mu^*_{\alpha,{\rm G,B}}-\hat\mu^*_{\alpha,{\rm G,B}})^2 C^{-1}_{\alpha\alpha,{\rm G,B}} \nonumber \\
& + (\mu_{\delta,{\rm G,B}}-\hat\mu_{\delta,{\rm G,B}})^2 C^{-1}_{\delta\delta,{\rm G,B}} \nonumber \\
& + 2(\mu^*_{\alpha,{\rm G,B}}-\hat\mu^*_{\alpha,{\rm G,B}})(\mu_{\delta,{\rm G,B}}-\hat\mu_{\delta,{\rm G,B}})
C^{-1}_{\alpha\delta,{\rm G,B}}, \nonumber 
\end{align}
in which the proper motions from \hip\ correspond to the primary. The
$C^{-1}$ terms are the relevant elements of the inverse of the
covariance matrix, which we computed from the uncertainties and
correlations given in Table~\ref{tab:pm}.

The model-predicted quantities $\hat\rho$ and $\hat\theta$ in the
expressions above are a function of the elements of the relative
orbit. The quantities $\hat\mu^*_{\alpha,{\rm H}}$,
$\hat\mu^*_{\alpha,{\rm G,A}}$, and $\hat\mu^*_{\alpha,{\rm G,B}}$, and
similar ones for the Declination components, were computed by adding a
perturbation to the p.m.\ of the barycenter calculated from the
temporal change in the position in the orbit of the relevant star. For
example, $\hat\mu^*_{\alpha,{\rm G,A}} = \mu^*_{\alpha,0} +
\Delta\mu^*_{\alpha,{\rm G,A}}$, in which the last term is the
derivative of the orbital position of the primary relative to the
barycenter at the time of the \gaia\ measurement as listed in
Table~\ref{tab:pm}. Even though \hip\ and \gaia\ observed
\vstar\ during a finite period of time (about 2.5 and 1.7 years,
respectively), those intervals are much shorter than the orbital
period, so we have considered the proper motions to be effectively
instantaneous.

In a similar fashion, we computed $\hat\mu^*_{\alpha,{\rm HG}}$ as
$$\hat\mu^*_{\alpha,{\rm HG}} = \mu^*_{\alpha,0} +
\frac{\Delta\alpha^*_{\rm G,A}[t_{\alpha,\rm G}] - \Delta\alpha^*_{\rm
    H}[t_{\alpha,\rm H}]}{t_{\alpha,\rm G} - t_{\alpha,\rm H}},$$
in which the $\Delta\alpha^*$ quantities represent the position of the
primary relative to the barycenter at the mean epoch of each catalog,
and are easily calculated from the orbital elements. A similar
equation was used for $\hat\mu_{\delta,{\rm HG}}$.

The MCMC solutions were repeated several times to manually fine-tune
the uncertainties $\sigma_{\rho}$ and $\sigma_{\rm t}$ that we
assigned to each of the four groups into which we separated the visual
observations (Section~\ref{sec:visual}). This was done to set the
relative weighting among groups. During each of those solutions we
simultaneously allowed the model to freely adjust the $f_{\rho}$ and
$f_{\theta}$ factors in order to balance the relative weight of the
angular separation and position angle measurements by setting the
absolute scale of their errors. This process led to the final
$\sigma_{\rho}$ and $\sigma_{\rm t}$ values given earlier in
Section~\ref{sec:visual}.

The results of our analysis are presented in Table~\ref{tab:mcmc},
with an indication of the priors used for each adjustable parameter.
The bottom section of the table gives the derived quantities computed
directly from the chains of adjusted elements, including the masses,
the linear semimajor axis, and periastron distance $a(1-e)$.  The
scaling needed to compute these derived quantities is set by the
parallax, which we assumed to be uncorrelated with the other
properties, and which is given by the weighted average of the values
in Table~\ref{tab:pm}. We list this average parallax in the table as
well.  In each case we report the mode of the posterior distributions,
along with the 68.3\% credible intervals.  Table~\ref{tab:cor} lists
the correlation coefficients among all adjustable parameters, with the
highest ones underlined.

\setlength{\tabcolsep}{4pt}
\begin{deluxetable}{lcc}
\tablewidth{0pc}
\tablecaption{Results of our MCMC Analysis for \vstar \label{tab:mcmc}}
\tablehead{ \colhead{~~~~~~~~~~~Parameter~~~~~~~~~~~} & \colhead{Value} & \colhead{Prior} }
\startdata
 $P$ (year)\dotfill               &  $172.5^{+2.0}_{-2.1}$  & [2, 7] \\ [1ex]
 $a$ (arcsec)\dotfill             &  $0.908^{+0.044}_{-0.017}$  & [$-$2, 2] \\ [1ex]
 $\sqrt{e}\cos\omega_{\rm B}$\dotfill  &  $-0.897^{+0.058}_{-0.025}$ & [$-$1, 1] \\ [1ex]
 $\sqrt{e}\sin\omega_{\rm B}$\dotfill  &  $0.34^{+0.13}_{-0.11}$ & [$-$1, 1] \\ [1ex]
 $\cos i$\dotfill                 &  $-0.432^{+0.035}_{-0.058}$     & [$-$1, 1] \\ [1ex]
 $\Omega_{2000}$ (degree)\dotfill &  $10.8^{+2.7}_{-4.7}$                 & [$-$180, 180] \\ [1ex]
 $T$ (year)\dotfill               &  $1891.9^{+1.5}_{-1.7}$ & [1800, 2000] \\ [1ex]
 $f \equiv M_{\rm B}/(M_{\rm A}+M_{\rm B})$\dotfill  &  $0.402^{+0.040}_{-0.043}$  &  [0, 1] \\ [1ex]
 $\mu^*_{\alpha,0}$ (mas yr$^{-1}$)\dotfill & $+106.41^{+0.23}_{-0.23}$  & [80, 120] \\ [1ex]
 $\mu_{\delta,0}$ (mas yr$^{-1}$)\dotfill & $-25.72^{+0.25}_{-0.27}$  & [$-$50, 0] \\ [1ex]
 $f_{\rho}$\dotfill                &  $1.487^{+0.087}_{-0.074}$  & [$-5$, 3] \\ [1ex]
 $f_{\theta}$\dotfill              &  $1.454^{+0.083}_{-0.071}$ & [$-5$, 3] \\ [1ex]
\noalign{\hrule} \\ [-1.5ex]
\multicolumn{3}{c}{Derived quantities} \\ [0.5ex]
\noalign{\hrule} \\ [-1.5ex]
 $e$\dotfill                      &  $0.915^{+0.020}_{-0.018}$  & \nodata \\ [1ex]
 $\omega_{\rm B}$ (degree)\dotfill &  $159.2^{+6.4}_{-8.6}$ & \nodata \\ [1ex]
 $i$ (degree)\dotfill             &  $115.5^{+3.8}_{-2.1}$ & \nodata \\ [1ex]
 $M_{\rm A}+M_{\rm B}$ ($M_{\sun}$)\dotfill  &  $2.72^{+0.46}_{-0.17}$ & \nodata \\ [1ex]
 $M_{\rm A}$ ($M_{\sun}$)\dotfill &  $1.63^{+0.30}_{-0.13}$ & \nodata \\ [1ex]
 $M_{\rm B}$ ($M_{\sun}$)\dotfill &  $1.11^{+0.21}_{-0.14}$ & \nodata \\ [1ex]
 $q \equiv M_{\rm B}/M_{\rm A}$\dotfill  &  $0.66^{+0.13}_{-0.10}$ & \nodata \\ [1ex]
 $\pi$ (mas)\dotfill              &  $20.984^{+0.060}_{-0.060}$  & \nodata \\ [1ex]
  Distance (pc)\dotfill           &  $47.66^{+0.14}_{-0.14}$   & \nodata  \\ [1ex]
 $a$ (au)\dotfill                 &  $43.26^{+2.09}_{-0.80}$ & \nodata \\ [1ex]
 $a(1-e)$ (au)\dotfill            &  $3.67^{+0.78}_{-0.79}$ & \nodata 

\enddata

\tablecomments{The values listed correspond to the mode of the
  respective posterior distributions, and the uncertainties represent
  the 68.3\% credible intervals. All priors are uniform over the
  specified ranges, except for $P$, $a$, $f_{\rho}$ and $f_{\theta}$,
  which are log-uniform. The parallax reported is the weighted average
  of the values listed in Table~\ref{tab:pm}.}

\end{deluxetable}
\setlength{\tabcolsep}{6pt}

\setlength{\tabcolsep}{5pt}  
\begin{deluxetable*}{lccccccccccccc}
\tablewidth{0pc}
\tablecaption{Correlation Coefficients Among the Adjusted Parameters of \vstar \label{tab:cor}}
\startdata
\hline\hline \\ [-1ex]
$\ln P$                       &  +1.000   & \nodata  & \nodata  & \nodata  & \nodata  & \nodata  & \nodata  & \nodata  & \nodata  & \nodata  & \nodata  & \nodata \\
$\ln a$                       &  $-$0.420 & +1.000   & \nodata  & \nodata  & \nodata  & \nodata  & \nodata  & \nodata  & \nodata  & \nodata  & \nodata  & \nodata \\
$\sqrt{e}\cos\omega_{\rm B}$  &  $-$0.504 & \underline{+0.984}   & +1.000   & \nodata  & \nodata  & \nodata  & \nodata  & \nodata  & \nodata  & \nodata  & \nodata  & \nodata \\
$\sqrt{e}\sin\omega_{\rm B}$  &  $-$0.639 & \underline{+0.906}   & \underline{+0.952}   & +1.000   & \nodata  & \nodata  & \nodata  & \nodata  & \nodata  & \nodata  & \nodata  & \nodata \\
$\cos i$                      &  +0.498   & +0.307   & +0.168   & +0.001   & +1.000   & \nodata  & \nodata  & \nodata  & \nodata  & \nodata  & \nodata  & \nodata \\
$\Omega_{2000}$               &  +0.742   & $-$0.788 & $-$0.876 & \underline{$-$0.948} & +0.289   & +1.000   & \nodata  & \nodata  & \nodata  & \nodata  & \nodata  & \nodata \\
$T$                           &  $-$0.316 & $-$0.631 & $-$0.547 & $-$0.451 & $-$0.747 & +0.193   & +1.000   & \nodata  & \nodata  & \nodata  & \nodata  & \nodata \\
$f$                           &  $-$0.158 & +0.038   & $-$0.016 & $-$0.032 & +0.195   & +0.095   & $-$0.072 & +1.000   & \nodata  & \nodata  & \nodata  & \nodata \\
$\mu^*_{\alpha,0}$            &  +0.151   & +0.004   & +0.053   & +0.067   & $-$0.132 & $-$0.110 & +0.035   & \underline{$-$0.978} & +1.000   & \nodata  & \nodata  & \nodata \\
$\mu_{\delta,0}$              &  +0.143   & +0.049   & +0.104   & +0.119   & $-$0.171 & $-$0.170 & +0.005   & \underline{$-$0.981} & \underline{+0.971}   & +1.000   & \nodata  & \nodata \\
$\ln f_{\rho}$                &  +0.051   & +0.013   & +0.008   & $-$0.027 & +0.055   & +0.034   & $-$0.010 & $-$0.028 & +0.033   & +0.032   & +1.000   & \nodata \\
$\ln f_{\theta}$              &  $-$0.065 & +0.040   & +0.043   & +0.051   & $-$0.061 & $-$0.064 & +0.013   & +0.012   & $-$0.016 & $-$0.013 & $-$0.019 & +1.000  \\ [1ex]
& $\ln P$ & $\ln a$ & $\sqrt{e}\cos\omega_{\rm B}$ & $\sqrt{e}\sin\omega_{\rm B}$ & $\cos i$ & $\Omega_{2000}$ & $T$ & $f$ & $\mu^*_{\alpha,0}$ & $\mu_{\delta,0}$ & $\ln f_{\rho}$ & $\ln f_{\theta}$
\enddata
\tablecomments{Correlation coefficients larger than 0.900 in absolute value are underlined.}
\end{deluxetable*}
\setlength{\tabcolsep}{6pt}  

\begin{figure}
\epsscale{1.15}
\plotone{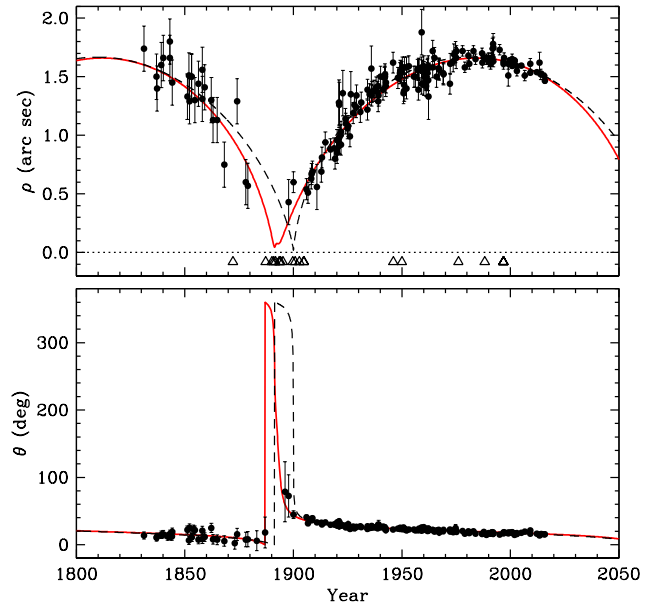}

\figcaption{Observations of \vstar\ along with our adopted model from
  Table~\ref{tab:mcmc}. The dashed line represents the orbit by
  \cite{Izmailov:2019}. Also shown in the top panel are the dates of
  the observations that did not resolve the binary
  (triangles). \label{fig:rhotheta}}

\end{figure}

\begin{figure}
\epsscale{1.15}
\plotone{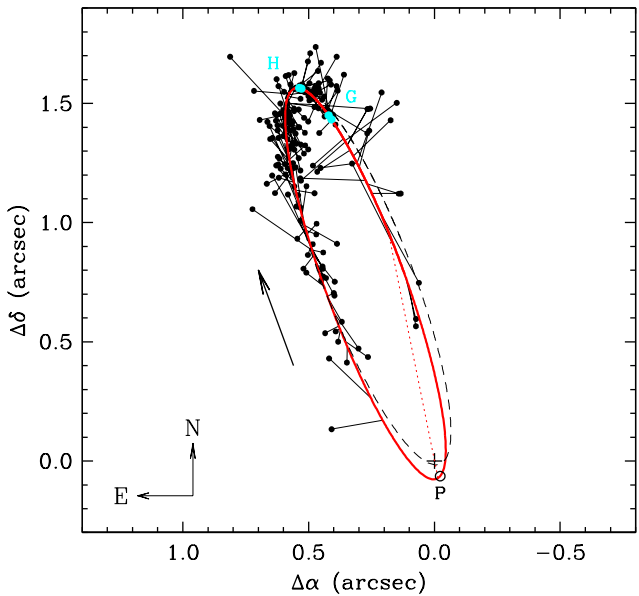}

\figcaption{Same as Figure~\ref{fig:rhotheta} in polar coordinates,
  again showing the orbit by \cite{Izmailov:2019} with a dashed line,
  for reference. The direction of motion (retrograde) is indicated by
  the arrow. Thin lines connect each observation with the predicted
  position in the orbit. The plus sign marks the position of the
  primary, and the dotted line is the line of nodes. Periastron is
  indicated with an open circle at the bottom, labeled ``P''. The
  epochs of the \hip\ and \gaia\ observations (1991.25 and 2015.5) are
  indicated with larger cyan filled circles near the
  top. \label{fig:polar}}

\end{figure}

The eccentricity, $e = 0.915^{+0.020}_{-0.018}$, is found to be less
extreme than the latest reported value of $0.979 \pm 0.011$ by
\cite{Izmailov:2019}.\footnote{We note that the study of
  \cite{Izmailov:2019} reported two orbital solutions for \vstar: one
  in which all observations were given equal weight, and another in
  which the measurements were weighted differently. Both were computed
  using a least-squares method followed by a Monte Carlo sampling
  experiment to infer uncertainties. The adopted weights for the
  second solution resulted from a complicated procedure that is not
  entirely clear to the present author, apparently involving an
  assessment of measurement residuals from different sources and also
  from other binaries, although no specific details were given for
  \vstar. The solution favored by \cite{Izmailov:2019} is the one that
  uses different weights, which gives the eccentricity we reported. In
  the other solution the eccentricity is slightly smaller ($e = 0.925
  \pm 0.019$). However, it is puzzling to see that the weighted
  solution has considerably larger uncertainties for most of the
  elements than the unweighted one, some of which are surprisingly
  large. The angular semimajor axis, for example, was reported as $a =
  1\farcs03 \pm 0\farcs48$, whereas the corresponding result from the
  unweighted orbit is $1\farcs02 \pm 0\farcs08$. Similarly with the
  orientation angles, which have uncertainties of tens of degrees in
  the weighted model, versus single digits for the unweighted one.  As
  a result, the implied total mass of the binary from the weighted
  solution (adopting the parallax in Table~\ref{tab:mcmc}) is $4.0 \pm
  8.4~M_{\sun}$, which is not only of little use but is also much
  worse than the value from the unweighted solution, $4.0 \pm
  1.0~M_{\sun}$. \label{foot:izmailov}} Our semimajor axis is somewhat
smaller compared to the estimates by both \cite{Baize:1980} and
\cite{Izmailov:2019}, but larger than that of \cite{Peterson:1988}.
Our adopted model is shown in Figure~\ref{fig:rhotheta} along with the
observations, with the orbit by \cite{Izmailov:2019} added for
reference (see footnote~\ref{foot:izmailov}). We find that our orbit
and that of \cite{Izmailov:2019} provide an equally good fit to the
observations after periastron, but that ours represents the earlier
measurements slightly better. Observations for which the observers
reported the binary to be unresolved are marked in the top panel of
Figure~\ref{fig:rhotheta} with triangles; they are seen to be mostly
at periastron, though a few have occurred at other times reflecting
the difficulty of detecting the faint secondary, or perhaps poor
observing conditions.  A rendering of the orbit in polar coordinates
is seen in Figure~\ref{fig:polar}. The predicted radial-velocity
difference between the components at the present time is only
2.1~\kms, and will increase very slowly to about 2.2~\kms\ by the year
2025 and 2.3~\kms\ by 2030.

\section{The Mass-Luminosity Relation in the Hyades}
\label{sec:MLR}

Separate brightness measurements for the components of the
\vstar\ binary are available from the Tycho-2 Catalogue
\citep{Hog:2000}. Conversion of those $B_{\rm T}$ and $V_{\rm T}$
magnitudes to the Johnson system gives $V$-band magnitudes of $5.674
\pm 0.010$ and $8.055 \pm 0.016$. Absolute visual magnitudes then
follow from the distance to the system (Table~\ref{tab:mcmc}), and are
$M_{\rm V}^{\rm A} = 2.283 \pm 0.012$ and $M_{\rm V}^{\rm B} = 4.664
\pm 0.017$, ignoring extinction. Although brightness measurements for
both components are also given in the \gaia/DR2 catalog, the values
for the secondary star in the blue and red bandpasses ($G_{\rm BP}$
and $G_{\rm RP}$) are very uncertain and potentially biased, according
to the accompanying quality flag, and we do not use them
here. Conversion of the flux measurements for the primary to the
Johnson system using the relation by \cite{Evans:2018} gives a value
consistent with the one above ($V = 5.641 \pm 0.046$).

These absolute magnitudes and our dynamical masses offer a valuable
opportunity to revisit the empirical MLR in the Hyades, especially
given that mass estimates (and in some cases the absolute magnitudes)
for some of the previously known systems have been revised in recent
years. A summary of those revisions is presented below.

$\bullet$~{\bf $\theta^2$\thinspace Tau.} The dynamical masses by
\cite{Tomkin:1995} were improved by \cite{Torres:1997c}, and were
supplemented with absolute magnitudes derived from the orbital
parallax. \cite{Armstrong:2006} made use of new interferometric
observations and the mass ratio from the previous study, and revised
the masses downward by about 0.25~$M_{\sun}$, also making the
uncertainties somewhat smaller. They replaced the orbital parallax of
the system with a dynamical parallax, but obtained very nearly the
same absolute magnitudes. Subsequently \cite{Lampens:2009} reported
new spectroscopic results from spectral disentangling that improved
the mass uncertainties by a factor of several, but made both masses
larger again. In the case of the primary ($2.58 \pm 0.04~M_{\sun}$),
which is a well known $\delta$\thinspace Sct star, the result is
formally larger than the expected turnoff mass of the cluster
($\sim$2.4~$M_{\sun}$). This is difficult to understand. A more recent
disentangling study by \cite{Torres:2011}\footnote{No relation to the
  present author.}  also resulted in small errors, and an even larger
primary mass ($2.86 \pm 0.06~M_{\sun}$) that seems implausible for a
main-sequence star in the Hyades \citep[see also the remarks
  by][]{Brandt:2015}. It is unclear what may be causing these larger
masses from the spectral disentangling technique.  Consequently, and
until those results can be understood, the determinations by
\cite{Armstrong:2006} seem preferable.

$\bullet$~{\bf 51\thinspace Tau.} The original mass determinations by
\cite{Torres:1997a} were revised by \cite{Martin:1998} using the
\hip\ astrometry combined with a relative astrometric orbit from
\cite{Balega:1988}, which resulted in a lower mass sum, a lower mass
ratio, but significantly larger uncertainties. \cite{Soderhjelm:1999}
also combined \hip\ observations with ground-based data, reporting
masses closer to those from the original determination though again
with larger uncertainties. The details of his solution were not
given. We have chosen to accept the \cite{Torres:1997a} results.

$\bullet$~{\bf 70\thinspace Tau.} The only published dynamical mass
determinations for this binary are those of \cite{Torres:1997b}, which
we adopt.

$\bullet$~{\bf $\theta^1$\thinspace Tau.} The original mass
determinations by \cite{Torres:1997c} were revised by
\cite{Lebreton:2001} using the relative orbit from the previous study
and a slightly larger dynamical parallax than that employed
earlier. The mass values are not very different, but the uncertainties
for the primary mass and for the absolute magnitudes are slightly
smaller, so we selected those results.

$\bullet$~{\bf V818\thinspace Tau.} The mass determinations of
\cite{Schiller:1987} were slightly updated by \cite{Peterson:1988}.
The latest estimates of the masses and absolute visual magnitudes of
this eclipsing system are those of \cite{Torres:2002}, which we use in
the following. The precision of the primary mass is slightly improved.

The adopted values for these five binaries (excluding the giant
primary of $\theta^1$\thinspace Tau) are collected in
Table~\ref{tab:MLR} in order of decreasing primary mass. Our own
determinations for \vstar\ are added for completeness. To our
knowledge these are the only binary systems in the Hyades with
dynamical mass determinations for the individual components. While
several other systems in the cluster have known visual orbits, they
can only provide a total mass (when combined, e.g., with parallaxes
from \gaia), and are therefore not useful for our purpose.

\setlength{\tabcolsep}{6pt}
\begin{deluxetable*}{lccccl}
\tablewidth{0pc}
\tablecaption{Dynamical Mass Determinations in the Hyades Cluster \label{tab:MLR}}
\tablehead{
\colhead{Binary System} &
\colhead{$M_{\rm A}$ ($M_{\sun}$)} &
\colhead{$M_{\rm B}$ ($M_{\sun}$)} &
\colhead{$M_V^{\rm A}$ (mag)} &
\colhead{$M_V^{\rm B}$ (mag)} &
\colhead{Source}
}
\startdata
$\theta^2$ Tau\dotfill & $2.15 \pm 0.12$     & $1.87 \pm 0.11$     & $0.48 \pm 0.08$ & $1.61 \pm 0.06$ & \cite{Armstrong:2006} \\
51 Tau\dotfill         & $1.80 \pm 0.13$     & $1.46 \pm 0.18$     & $2.14 \pm 0.10$ & $3.75 \pm 0.17$ & \cite{Torres:1997a} \\
80 Tau\dotfill         & $1.63^{+0.30}_{-0.13}$  & $1.11^{+0.21}_{-0.14}$   & $2.283 \pm 0.012$ & $4.664 \pm 0.017$ & This work \\ [0.5ex]
70 Tau\dotfill         & $1.363 \pm 0.073$   & $1.253 \pm 0.075$   & $3.71 \pm 0.10$ & $4.05 \pm 0.11$ & \cite{Torres:1997b} \\
$\theta^1$ Tau\dotfill & \nodata             & $1.28 \pm 0.13$     & \nodata         & $4.02 \pm 0.07$ & \cite{Lebreton:2001} \\
V818 Tau\dotfill       & $1.0591 \pm 0.0062$ & $0.7605 \pm 0.0062$ & $4.93 \pm 0.09$ & $7.23 \pm 0.12$ & \cite{Torres:2002} 
\enddata
\end{deluxetable*}
\setlength{\tabcolsep}{6pt}

\subsection{Comparison with Models}
\label{sec:models}

\begin{figure}
\epsscale{1.15}
\plotone{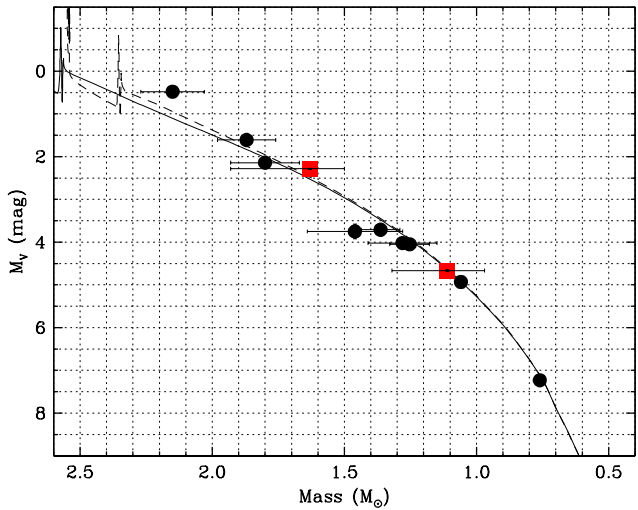}

\figcaption{Empirical mass-luminosity relation in the Hyades based on
  the six binary systems in the cluster with dynamical mass
  determinations for the individual components (Table~\ref{tab:MLR}).
  Square symbols represent the estimates for \vstar\ from this
  work. Isochrones from the PARSEC series \citep{Chen:2014} are shown
  for ages of 625~Myr (solid line) and 800~Myr (dashed), for a fixed
  cluster metallicity of ${\rm [Fe/H]} = +0.18$.\label{fig:MLR}}

\end{figure}

The empirical mass-luminosity relation based on the data in
Table~\ref{tab:MLR} is presented in Figure~\ref{fig:MLR}.
Measurements for \vstar\ are represented with square symbols. Stellar
evolution models from the PARSEC series by \cite{Chen:2014}
corresponding to a metallicity of ${\rm [Fe/H]} = +0.18$ are shown as
well, for ages of 625~Myr \citep[e.g.,][]{Perryman:1998} and also
800~Myr \citep{Brandt:2015}. Other model isochrones such as those from
the MIST series \citep{Choi:2016} are almost indistinguishable in this
diagram, and are not shown. The agreement between theory and
observation is quite satisfactory, with nearly all mass estimates
being consistent with the isochrones within the (admittedly large)
observational errors, some of which exceed 10\%.  The highly
asymmetric error bars for the masses of \vstar\ are a consequence of
the shape of the posterior distributions.

While our choices above regarding the sources for the dynamical masses
of the five previously studied binaries have generally been driven by
the level of precision (formal errors), this is not necessarily a
guarantee of accuracy. We note, for example, that adopting the masses
of \cite{Soderhjelm:1999} for 51\thinspace Tau instead of those of
\cite{Torres:1997a} would give a virtually perfect match to the
models, though with larger error bars. Similarly, the more uncertain
masses of \cite{Torres:1997c} for $\theta^2$\thinspace Tau agree
better with theory than those of \cite{Armstrong:2006}. Both of these
may well be accidental agreements. In any case, an underlying weakness
of many if not all of the Hyades mass determinations is the fact that
they have often used partial results from earlier studies such as
fixed visual orbital elements, fixed velocity semi-amplitudes, fixed
mass ratios, etc., rather than the original data on which those
external constraints were based. This is not optimal. In some cases
this was done because the original data were not available. 
  Although it is beyond the scope of the present study, a reanalysis
of several of these systems would be beneficial, and could improve not
only the precision but also the accuracy of the mass determinations.
For some of the binaries more than 20 years have passed since the last
detailed analysis (which is longer than a full orbital cycle even for
$\theta^1$\thinspace Tau, with $P = 16.3$~yr), and new and better
observations may be available.

\section{Concluding remarks}
\label{sec:remarks}

We have measured the absolute masses of the components of the Hyades
visual binary \vstar, only the sixth system with individual dynamical
mass determinations in the cluster, and the first to be identified in
more than 20 years. It is also the first case that does not rely on
radial-velocity observations, which are very challenging here because
of the long orbital period ($\sim$170~yr) and the rapid rotation of
the primary star. Instead, we have succeeded in deriving the masses
using only astrometric observations. The key ingredients supplementing
the visual measurements of the pair are the individual proper motions
measured by \gaia\ for the two components, which, combined with the
p.m.\ for the primary from the \hip\ mission, provide the only
constraint available on the mass ratio. We infer an orbit that is
highly inclined and very eccentric ($e = 0.915$), such that at closest
approach the two stars come within about 3.7~au of each other.

A vexing problem in the use of visual observations for orbit
computation is the assignment of realistic uncertainties, which are
rarely reported for the historical measurements. While sensible, the
weighting scheme adopted for those measurements in this work is by no
means the only one possible. We note, for example that an earlier
analysis we carried out leading up to the present one differed in
three respects regarding the treatment of errors: 1) the observations
were separated by date into three groups instead of four, giving lower
weight to the speckle measurements than we do now; 2) constant errors
in degrees were adopted for the position angles within each group,
instead of constant errors in seconds of arc in the tangential
direction, as we have done now to account for the error dependence on
angular separation; and 3) no adjustment of the overall scale of the
errors was made to balance the position angle and separation
residuals, as done in the present work through the free parameters
$f_{\rho}$ and $f_{\theta}$. Reassuringly, despite these differences
the mass estimates turn out to be very similar between the two
procedures, suggesting little dependence on the details of how errors
are assigned, at least in this particular case.

Our mass determinations along with those of the five previously known
systems define an empirical MLR for the cluster in good agreement with
current models of stellar evolution, given present uncertainties. We
note, however, that several of the mass measurements ---including our
own--- still have relative errors exceeding 10\%, and some of those
determinations are now quite dated. New observations of higher quality
may well be available in many cases. It seems likely that some of the
masses could be improved considerably in both accuracy and precision
with a self-consistent reanalysis of all available observations that
avoids using partial results from earlier studies of these binaries.
This would allow for a more stringent comparison with stellar
evolution models than possible at this time.

\begin{acknowledgements}

We thank Robert Stefanik for bringing this system to our attention.
We also thank Brian Mason for providing a listing of the measurements
of \vstar\ from the Washington Double Star Catalog, and the anonymous
referee for helpful comments. This work has been supported in part by
grant AST-1509375 from the National Science Foundation. The research
has made use of the SIMBAD and VizieR databases, operated at the CDS,
Strasbourg, France, of NASA's Astrophysics Data System Abstract
Service, and of the Washington Double Star Catalog maintained at the
U.S.\ Naval Observatory. The work has also made use of data from the
European Space Agency (ESA) mission
\gaia\ (\url{https://www.cosmos.esa.int/gaia}), processed by the
\gaia\ Data Processing and Analysis Consortium (DPAC,
\url{https://www.cosmos.esa.int/web/gaia/dpac/consortium}). Funding
for the DPAC has been provided by national institutions, in particular
the institutions participating in the \gaia\ Multilateral Agreement.

\end{acknowledgements}

\end{document}